\documentstyle[12pt]{article}
\def\eg{{\it e.g.\ }} 
\def\etal{{\it et al.\ }} 
 
\def\ie{{\it i.e.\ }} 

\def\sb{{mag arcsec$^{-2}$\ }}
\def\solar{\ifmmode_{\mathord\odot}\else$_{\mathord\odot}$\fi}
\def\app{$\approx$~}
\begin{document}
\leftline{\LARGE {\bf HIDDEN GALAXIES REVEALED}}
\bigskip
\bigskip
\bigskip
\bigskip
\bigskip
\leftline{\Large {\sl Greg Bothun}}
\smallskip
\leftline{Physics Department, University of Oregon, Eugene, Oregon 97403}
\bigskip
\bigskip

\leftline{\Large {\sl Chris Impey}}
\smallskip
\leftline{Steward Observatory, University of Arizona, Tucson, Arizona 85721}
\bigskip
\bigskip
\leftline{\Large {\sl Stacy McGaugh}}
\smallskip
\leftline{Department of Terrestrial Magnetism, Carnegie Institute of Washington}
\bigskip
\bigskip
\bigskip
\leftline{KEY WORDS:\quad Galaxies, physical properties, galaxy morphology, 
cosmology}
\bigskip
\bigskip
An Invited Review Article for Publications of the Astronomical Society
of the Pacific.
\clearpage
\centerline{\small ABSTRACT}
\bigskip
\bigskip

In twenty years, low surface brightness (LSB) galaxies have evolved from
being an idiosyncratic notion to being one of the major baryonic
repositories in the Universe.  The story of their discovery and
the characterization of their properties is told here.  Their recovery
from the noise of the night sky background
is a strong testament to the severity of surface brightness selection
effects.  LSB galaxies have a number of remarkable properties which
distinguish them from the more familiar Hubble Sequence of spirals.
The two most important are 
1) they evolve at a significantly slower rate
and may well experience star formation outside of the molecular
cloud environment, 2) 
they are embedded in dark matter halos which are of lower
density and more extended than the halos around high surface brightness (HSB)
disk galaxies.   Compared to HSB disks, LSB disks are strongly
dark matter dominated at all radii and show a systematic increase in
$M/L$ with central surface brightness.

\bigskip

In addition, the recognition that large numbers of LSB
galaxies actually exist has changed the form of the galaxy luminosity
function and has clearly increased the space density of galaxies
at z =0.  Recent CCD surveys have uncovered a population 
of red LSB disks that may be related to the excess of
faint blue galaxies detected
at moderate redshifts.
LSB galaxies offer us a new window into galaxy evolution and
formation which is every bit as important as those processes which
have produced easy to detect galaxies.   Indeed, the apparent youth
of some LSB galaxies suggest that galaxy formation is a greatly
extended process.  While the discovery of LSB galaxies have lead
to new insights, it remains unwise to presume that we now
have a representative sample which encompasses all galaxy types and
forms.  
\vfil\eject

\leftline{\large 1.\quad INTRODUCTION}
\bigskip

Through the noisy haze of sky photons, astronomers since the time of
Messier have detected and cataloged the positions and shapes of  diffuse,
resolved objects known as  nebulae.   The cataloger well knows the limits
on sensitivity posed by the observing environment, yet these limits
are rarely quantified and passed on to the next generation of
astronomers.  Indeed, if Messier were alive in today's light polluted
world, his catalog would certainly be much more sparse since he could
only catalog the nebulae he could see.   Given this basic constraint,
the natural question to ask is ``Are there diffuse nebulae that cannot 
be cataloged because they remain masked by the night sky?''   For the 
case of galaxy detection, this question is quite relevant in the context 
of the Cosmological Principle, a corollary of which asserts that all 
observers in the universe should construct similar catalogs of galaxies. 
If this were not the case, then different observers might have biased views
and information about (1) the nature of the general galaxy population in
the Universe, (2) the three dimensional distribution of galaxies, and
(3) the amount of baryonic matter that is contained in galactic potentials.
On the largest scales, we expect the universe to exhibit a homogeneous
appearance, but our only signposts for matter are the galaxies 
whose light we detect with optical telescopes against a noisy background
of finite brightness.
Given this condition one can easily conceive of observing 
environments that would make galaxy detection difficult.

\bigskip

For example, suppose that we lived on a planet that was located in the
inner regions  of an elliptical galaxy.  The high stellar density would
produce a night sky background that would be relatively bright and
therefore not conducive to the discovery of galaxies.  Similarly, if the
Solar System in its journey around the galaxy were unlucky enough to
be located near or in a Giant Molecular Cloud (GMC) at the same time that
evolutionary processes produced telescopes on the Earth, then our
observational horizon would be severely limited by the local dust associated
with the GMC.  As it is, we are fortunate enough to be located at a 
relatively dust free area $\sim$ 2.5 scale lengths from the center of 
a spiral galaxy.  At this distance, the local surface brightness of the 
galactic disk is $\sim$ 24 \sb in the blue which gives us a relatively dark 
window to peer out towards the galactic poles and discover diffuse objects.
Does this relatively unobscured view guarantee that earth-bound extragalactic
astronomers are able to detect a representative sample of galaxies?  

\bigskip

The idea that the night sky emission places limits on the kinds
of galaxies which can be detected was first commented on by Zwicky (1957).
The first quantitative analysis of the potential magnitude of this
selection effect was presented by Disney (1976).   Disney's efforts
were largely motivated by the discovery of Freeman (1970) that
spiral galaxies seemed to exhibit a constant central surface
brightness ($\mu_0$) in the blue.  The formal value found by Freeman was
$\mu_0$ = 21.65 $\pm$ 0.35 for a sample of a few dozen spirals.
This constancy of $\mu_0$ became known as ``Freeman's Law.'' Like
any law, it was apparently made to be broken.  This review, 20 years 
after Disney's original analysis, shows that his basic argument
has been vindicated.  Selection effects have been severe and as a result 
{\it no representative sample of nearby galaxies has yet been compiled, 
cataloged and investigated.} 

\bigskip

The most dramatic confirmation that these selection effects are real and 
significant has been provided by McGaugh \etal (1995) and is reproduced 
here in Figure 1.  Ten years of hunting for galaxies of
low surface brightness (LSB) has revealed
a surprising result which subverts the conventional wisdom  as
embodied by Freeman's Law.   
Figure 1 shows that up to 50\% of the general population of galaxies 
resides in a continuous tail extending towards low 
$\mu_0$.  Thus, the space density of LSBs is significant.  This conclusion 
has also been reached by Dalcanton \etal (1997) from a study of 7 LSB 
galaxies detected in the Palomar 5-m transit scan data.  With measured
redshifts they assigned a tentative space density of $0.02^{+0.02}_{-0.01} 
h_{50}^3 {\rm Mpc}^{-3}$ for galaxies with $\mu_0$ fainter than 23.5 \sb.

\bigskip

As emphasized by McGaugh \etal (1995), the most physically reasonable 
approach in converting raw counts to space densities is
to assume that scale length and absolute galaxy magnitude are
uncorrelated.  As shown explicitly below, this results in smaller volumes
being accessible to surveys for LSB galaxies
compared to surveys for ``normal'' or high surface brightness (HSB) galaxies.
It is HSB galaxies that define the Hubble Sequence from which Freeman
(1970) derived his sample.  The space distribution of galaxies as 
a function of $\mu_o$ after the volume sampling correction has been
applied produces the distribution shown in Figure 1.  The dark point
defined by the Schombert \etal (1992) survey has a space density which
is 10$^5$ times higher than the extrapolation of Freeman's Law
would predict.  Factors of 10$^5$ are significant.  The space density
derived by Dalcanton \etal (1997) is even higher than this, perhaps suggesting
that galaxies become smaller at lower surface brightness.  The opposite
trend is seen in other data (\ie LSB galaxies tend if anything to be larger;
de Jong 1996); this illustrates the enormous uncertainty that remains in
our knowledge of the local galaxy population.  Nevertheless, the implication
of Figure 1 is clear -- very diffuse galaxies exist and they exist in large 
numbers.  Their properties are only now being elucidated. 

\bigskip
This review deals primarily with the physical properties of these newly 
discovered galaxies and their connection to galaxy evolution. For distance
dependent quantities, we assume $H_0$ = 100 kms$^{-1}$ Mpc$^{-1}$, and 
scale by $h_{100} = (H_0/100)$.  A companion review (Impey and Bothun 1997) 
more fully details the selection effects that have previously prevented 
the discovery of LSBs, and how their actual discovery impacts the proper
determination of the galaxy luminosity function and its relation to QSO 
absorption lines as well as deep galaxy surveys that have revealed an 
apparent excess of intermediate luminosity galaxies at intermediate 
redshifts.  The overall context of this review is the idea that LSB disk 
galaxies represent a parallel track of galaxy evolution that is largely 
decoupled from the processes that have determined the Hubble sequence.  
While this has significant implications for galaxy formation scenarios and
galaxy evolution  the existence of LSB disks themselves is not
a surprise. The interesting issue is epoch at which
they begin to appear and proliferate.  
Eventually, when astration is complete in disks, 
the universe will contain nothing but LSB objects.

\bigskip
\bigskip
\bigskip
\leftline{\large 2.\quad SURFACE BRIGHTNESS SELECTION EFFECTS}
\bigskip
\leftline{2.1 \quad How Surface Brightness Is Measured}
\bigskip

In simple terms, surface brightness is very similar to surface air 
pressure.  The amount of air molecules in a three dimensional column
of air in the atmosphere determines the total amount of pressure which
is exerted at a point on the {\it surface} of the earth.  If we
imagine a disk galaxy as an optically thin cylinder, then the surface
brightness is a measure of the space density of stars as projected
through a cylindrical cross section.   The mean luminosity density
through this cylinder, which is determined both by the stellar luminosity 
function and the mean separation between stars, is what observers
measure as a projected surface brightness.  Since the number of stars 
per Mpc$^3$ has a strong radial dependence, the projected surface brightness 
profile shows a fall off with radius.  This fall off is generally exponential 
in character and can be expressed as

$$ \mu(r) = \mu_0 + 1.086 {r\over\alpha_l}, $$

\noindent  Two parameters completely characterize the light distribution:
$\mu_0$ is the central surface light intensity and $\alpha_l$ is the scale 
length of the exponential light fall off.  In what follows $\mu_0$ will refer
to the central surface brightness in the blue.  If Freeman's law is  correct,
the number of parameters relevant to galaxy selection reduces to
one as variations in size modulate those in luminosity.
Since $\mu_0$ is a measure of the characteristic surface mass density
of a disk, Freeman's Law requires that all the physical processes of
galaxy formation and evolution conspire to result in this very specific
value for all galaxies.  Either the surface mass density must be
the same for all galaxies (in itself a peculiar result) with little
variation in the mass to light ratio, or
variations in the star formation history, collapse
epoch and initial angular momentum content must all conspire to balance
at this arbitrary value.

\bigskip

Because of the importance that Freeman's Law has with respect to
the physics of galaxy formation, there were several attempts to
explain it away.  While Disney (1976) dismissed Freeman's Law as an 
artifact of selection, others were not so sure.   For instance,
Kormendy (1977) asserted that Freeman's Law could be an artifact of improper
subtraction of the bulge component in disk galaxies.  Boroson (1981)
suggested that Freeman's Law was a conspiracy of dust obscuration, since
if galaxies had appreciable optical depth then we could only see their 
front surfaces. This could result in the appearance that disk surface 
brightnesses were fairly constant.  Bothun (1981) discovered
from a thorough survey of disk galaxies in the Pegasus I cluster that
there were equal numbers of galaxies with $B$ band central surface
brightness values in the range of 21-23.5 \sb.  The existence of any disk 
galaxy with $\mu_0$ fainter than 23.0 \sb would represent a 4$\sigma$ 
deviation from Freeman's Law and hence should not have been found so 
easily.  Unfortunately, Bothun (1981) was not sufficiently astute to 
notice or appreciate the significance of this at that time so the effect 
went unnoticed, until recently (\eg Figure 1).

\bigskip
In principle, the surface brightness profiles of galaxies can be traced
to arbitrarily large radii.  The only requirement being that those
stars are gravitationally bound to the galaxy.  Recent data by
Zaritsky (1997) shows that the dark matter halos of typical
spiral galaxies may be extremely large (up to 200 h$_{100}^{-1}$ kpc in
radius) and hence disk galaxies could extend out to \app 50 scale lengths!
Integration of equation 1, as a function of scale length, shows that
1$\alpha_l$ contains 26\% of the total luminosity, while 4$\alpha_l$ and
5$\alpha_l$ contain 90 and 96\%, respectively.  In practical terms, a 
diameter defined by four scale lengths provides a good measure of the 
total luminosity of the system.  

\bigskip

For a Freeman disk, this corresponds to an isophotal level of
26.05 \sb.  The darkest night skies that can be found for terrestrial
observing have $\mu_0 \sim$ 23.0 \sb.  Thus, the 90\% luminosity isophote
is some 3 mags below even the darkest skies or 6\% of the sky level.
We have consistently defined a LSB disk as one which has  $\mu_0$
fainter than 23.0 \sb or a 90\% luminosity isophote of 27.4 \sb
which is 2\% of the night sky brightness.
The Freeman value for  $\mu_0$ is about 1~magnitude brighter than the
surface brightness of the darkest night sky.  That the number of
galaxies with faint central surface brightnesses appears to decline
rapidly as $\mu_0 \rightarrow \mu_{\rm sky}$ is suspicious and if true 
of the real galaxy population implies that our observational viewpoint
is privileged in that we are capable of detecting most of the galaxies
that exist, at least when the moon is down.  This is the essence of the 
argument voiced by Disney (1976) in characterizing the Freeman Law as a 
selection effect.

\bigskip

So if surface brightness selection effects are important, what is the
best way to measure surface brightness?  Consider the following two
hypothetical galaxies which have the same total luminosity ($M_B = -20$):

\bigskip

$\bullet$ Galaxy A: \quad $\mu_0$ = 21.0 $\alpha_l$ = 3 kpc.

$\bullet$ Galaxy B: \quad $\mu_0$ = 24.0 $\alpha_l$ = 19 kpc.

\bigskip

\noindent There are three conceivable ways of measuring the surface
brightness in these disks: (1) central surface brightness ($\mu_0$), 
(2) average surface brightness within a standard isophote (= 25.0 \sb, 
$\mu_{\rm iso}$), or (3) effective surface brightness (measured within 
the half light radius = 1.7 $\alpha_l$ = $\mu_{\rm eff}$). Note that methods 
2 and 3 do not require that the light profile is well fit by an exponential 
function.  The results of applying each of these three measures of surface 
brightness are the following:

\bigskip

$\bullet$ Galaxy A: \quad $\mu_0$ = 21.0; $\mu_{\rm iso}$ = 23.2; 
$\mu_{\rm eff}$ = 22.3.

$\bullet$ Galaxy B: \quad $\mu_0$ = 24.0; $\mu_{\rm iso}$ = 25.8; 
$\mu_{\rm eff}$ = 24.8.

\bigskip

\noindent The largest difference in surface brightness occurs when 
$\mu_0$ is used as the measure.  Hence we have adopted this to define
the disk galaxy surface brightness.  While we realize that this definition 
requires that the galaxy be adequately fit by an exponential profile, most
LSB galaxies at all luminosities meet this criteria (see McGaugh and Bothun 
1994; O'Neil \etal 1997a; Sprayberry \etal 1995a).  This exercise also makes 
the trivial point that galaxies with low $\mu_0$ require surface photometry 
out to very faint isophotal radii in order to determine a total luminosity.

\bigskip
\leftline{2.2 \quad A Censored View of the Galaxy Population}
\bigskip

Much of  our knowledge of galaxies has stemmed from detailed studies
of objects that populate the Hubble Sequence.  Over 70 years ago, Hubble
(1922) warned of relying too much on this venture:

\bigskip

\indent
\lq\lq Subdivision of non-galactic nebulae is a much more difficult problem.
At present and for many years to come, their classification must rest solely
upon the simple inspection of photographic images, and will be confused, by
the use of telescopes of widely differing scales and resolving powers.
Whatever selection of types is made, longer exposures and higher resolving
powers will surely cause a reclassification of many individual nebulae ..."

\bigskip

\noindent In this quote Hubble establishes that galaxy classification, and 
therefore implicitly galaxy detection, is highly dependent upon observing 
equipment and resolution.  The essential issues are: (1) how severe is the 
bias in terms of the potential component of the galaxy population that has 
been missed to date, and (2) how would this effect our current understanding 
of galaxy formation and evolution?   In hindsight it is somewhat
mysterious why this issue of galaxy detection wasn't considered more
seriously 25 years ago.  Tinsley's elegant and accurate modeling of
the stellar populations of galaxies in the late 60's and early 70's
certainly indicated that galaxies could undergo significant luminosity
evolution, thereby producing faded and diffuse galaxies at z = 0.
Alternatively, there might be a population of intrinsically low surface 
mass density systems whose evolution is quite different from ``normal'' 
galaxies, but which nevertheless are important repositories of baryonic 
matter.   Disney was the most resonant voice to suggest that such diffuse 
systems could exist, and therefore that we could be missing an important 
constituent of the general galaxy population.

\bigskip

A very simple way of describing the effect of surface brightness 
selection is offered below.  While Disney and Phillips (1983) and
McGaugh \etal (1995) have quantified
these effects to produce Figure 1, the essential point is that LSB
disks, at any luminosity/circular velocity, are detectable out to
a significantly smaller distance than HSB disks.  Consider the
five hypothetical galaxies listed below.  The first four have pure
exponential light distributions and similar total luminosity
($M_B$ \app $-21.1$); the fifth galaxy has the same scale length has
Galaxy B but a factor of 10 lower total luminosity. 
This adheres to the McGaugh \etal
(1995) assumption that scale length and total luminosity are uncorrelated
in a representative sample of disk galaxies.

\bigskip

$\bullet$ Galaxy A: \quad $\alpha_l$ = 0.5 kpc, $\mu_0$ = 16.0 mag arcsec$^{-2}$

$\bullet$ Galaxy B: \quad $\alpha_l$ = 5.0 kpc, $\mu_0$ = 21.0 mag arcsec$^{-2}$

$\bullet$ Galaxy C: \quad $\alpha_l$ = 25.0 kpc, $\mu_0$ = 24.5 mag arcsec$^{-2}$

$\bullet$ Galaxy D: \quad $\alpha_l$ = 50.0 kpc, $\mu_0$ = 26.0 mag arcsec$^{-2}$

$\bullet$ Galaxy E: \quad $\alpha_l$ = 5.0 kpc, $\mu_0$ = 23.5 mag arcsec$^{-2}$

\bigskip

\noindent Assume that the intrinsic space density of these five galaxies are
equal, and suppose that we conduct a survey to catalog galaxies which have
diameters measured at the $\mu_0$ = 25.0 mag arcsec$^{-2}$ level ($D_{25}$)
of greater than one arcminute.  Under these conditions, we are interested in
determining the maximum distance that each galaxy can be detected:

\bigskip

Galaxy A: \quad This galaxy is quite compact (ratio of 1/2 light diameter 
to $D_{25}$ = 0.33) and would fall below the catalog limit beyond a distance
of 60 Mpc.

\bigskip

Galaxy B: \quad This is a typical large spiral (like M31); $D_{25}$ 
corresponds to 3.74 $\alpha_l$ which is 18.7 kpc or a diameter of 37 kpc.  
This projects to an angular size of 1 arcminute at a distance of 125 Mpc.

\bigskip

Galaxy C: \quad $D_{25}$ corresponds to 0.45 $\alpha_l$ or 11.5 kpc.  
This projects to an angular diameter of 1 arcminute at a distance of 76 Mpc.

\bigskip

Galaxy D: \quad $D_{25}$ doesn't exist and this galaxy would {\it never} 
be discovered in such a survey.

\bigskip

Galaxy E: \quad $D_{25}$ corresponds to 0.92 $\alpha_l$ or 4.6 kpc.   
This projects to an angular diameter of 1 arcminute at a distance of 30 Mpc.

\bigskip

The total survey volume is defined by Galaxy B, as they can be seen to
the largest distance.  The ratio of sampled volumes for each of the
other Galaxy types is considerably smaller.  For instance, the volume
ratio of Galaxy B to Galaxy E is a factor of 70!  Hence, a survey like
this would take the real space density distribution (which is equal) and, 
through the survey selection effect, produce a catalog which would contain 
72\% type B galaxies, 18\% type C galaxies, 9\% type A galaxies and 1\%
type E galaxies.  Type D galaxies would not be represented at all.
This is a severe bias which
would lead us to erroneously conclude that there is predominately one type 
of disk galaxy in the universe, and that this naturally leads to Freeman's Law.  
Now in the real zoology of LSB galaxies, type C and D galaxies are quite rare 
but type E galaxies are common.  Hence, their detection locally automatically
means the space density is relatively large because they are so heavily
selected against and the volume correction factors are significant.  This was 
the essence of Disney's original argument but there was no real data to 
support it at that time.  Twenty years of progress enables us to plot
Figure 1, which has the remarkable property that there is no indication 
of a rapid fall of in $\mu_0$. 

\bigskip
\bigskip
\bigskip
\leftline{\large 3.\quad SEARCH AND DISCOVERY}
\bigskip

The story of the discovery and characterization of LSB galaxies as important
members of the general galaxy population began in 1963 with the publication 
of the David Dominion Observatory (DDO) catalog of galaxies by Sydney van 
den Bergh.  This catalog consists of galaxies which exhibit a diffuse 
appearance and that  have angular sizes larger than three arc minutes.  
While the DDO objects are the first bona fide collection of a sample of 
LSB galaxies, they are {\bf not} at all representative of the phenomenon.  The 
galaxies contained in the DDO catalog are exclusively of low mass (some 
are members of the Local Group).  This has fostered an erroneous perception
that all LSB galaxies are dwarf galaxies. Today, we know that all masses of 
galaxies have representation in the LSB class.   

\bigskip

The first substantial contribution to our understanding of LSB disk
galaxies was made 
by William Romanishin and his collaborators Steve and Karen Strom in 1983 
(see Romanishin \etal 1983).  They derived their sample from the 
Uppsala General Catalog of Galaxies (UGC - Nilson 1973).  The UGC is 
a diameter selected catalog and is therefore less sensitive to surface
brightness selection effects compared to galaxy selection based on
apparent flux (see also McGaugh \etal 1995).  As a result, the UGC does
contain some LSB disk galaxies but most have  
$\mu_0$ in the range 22-23.0 \sb.  In early 1984, Alan Sandage and his
collaborators published some of the first results of the Las Campanas 
Photographic Survey of the Virgo Cluster.  Contained in those papers were 
some dramatic examples of dwarf galaxies in the Virgo cluster which were 
quite diffuse.  The existence of such diffuse objects in a cluster was very
interesting as this environment should prove hostile to their formation and 
survival.   This immediately raised the possibility that perhaps galaxies 
like this were common but uncataloged.  If so, perhaps these faint diffuse
galaxies were the source of the enigmatic QSO absorption line systems.
These considerations forged the Impey/Bothun collaboration as we collectively
wondered if the Sandage survey had missed galaxies of even lower surface
brightness. To answer this question we enlisted the help of David Malin in 
Australia. Ultimately, we were trying to improve the determination of the 
galaxy luminosity function by concentrating on those galaxies which would 
be the most difficult to detect, due to extreme diffuseness.  We had no idea 
if such extreme LSB galaxies really existed; to say we knew what we were doing 
would really exploit the advantage of hindsight!

\bigskip

Malin's method of photographic amplification had been used to find low surface
brightness shells and other tidal debris around normal galaxies, and it
could be extended to find entire very LSB galaxies.  Indeed,
Malin already had anecdotal evidence that whenever he Malinized a plate,
he would find these ``faint little buggers'' everywhere.  So David agreed
to Malinize selected one square degree areas of the Virgo cluster from
which several small diffuse objects emerged.  Skeptical
colleagues insisted that the peculiar collection of faint smudges that could 
be seen were all artifacts of the processing, water spots, or specks of dust.
Even the detection of 21-cm emission from one of these ``plate flaws'' did 
not assuage one particularly recalcitrant referee.  With youth, low pay
and foolishness on our side,
we persisted in our efforts to verify the reality of the smudge galaxies.

\bigskip

In late 1985 and early 1986 we used the Las Campanas 100-inch telescope for 
CCD imaging the very diffuse galaxies found in the Malinization process 
(all of which turned out to be real).  Most of these galaxies were devoid
of structure.  However, one had what appeared to be very faint spiral 
structure which was connected to a point-like nuclear region.  On the 
Palomar Sky Survey, this nuclear region is unresolved with no associated 
nebulosity apparent.  This was one of the few Malin objects bright enough
for optical spectroscopy, and on May 1986 at the Palomar 200-inch telescope, 
Jeremy Mould and Bothun took a spectrum of its nucleus.   Astonishingly, the 
spectrum exhibited emission lines at a redshift of z = 0.083, or a recessional
velocity of about 25,000 kms$^{-1}$.  Now, we had pursued the Malinization 
process on UK Schmidt plates of the Virgo cluster in order to find extremely 
LSB galaxies in the cluster.  Virgo has a mean recessional velocity of
1150 kms$^{-1}$, so this nucleated object clearly was far beyond Virgo.
Since the total angular size of the object on our CCD frame was approximately 
2.5 arcminutes, quick scaling then indicated that if a galaxy like this was 
indeed in Virgo then its angular size would be a degree. If it were as close 
as the Andromeda Galaxy its angular size would be about 20 degrees and of 
course we would look right through it without noticing it.  This seemed 
absurd, and there was a good chance that this strange galaxy was a 
composite system, consisting of a background emission line galaxy shining
through a foreground dwarf.

\bigskip

In October 1986, 21-cm observations at Arecibo revealed the characteristic 
signature of  a rotating disk galaxy whose systemic velocity was equal to 
that of the emission line object.  The accidental discovery of Malin 1 
(Bothun \etal 1987) strongly confirmed Disney's original speculation of
the existence of ``crouching giants.''   The existence of Malin 1 (type 
D in the previous example) certainly implied a non negligible space density 
of these kinds of objects.  The properties of Malin 1 are described in detail
by Impey and Bothun (1989). Recent H I observations of Malin 1 using the VLA
by Pickering \etal (1997) confirm the presence of a greatly extended gaseous 
disk around the normal bulge component of the galaxy.  

\bigskip

The remaining ``smudges'' in the Virgo cluster area did not turn out
to be as spectacular as Malin 1.  These smudges were most likely
LSB dwarf galaxy members of the Virgo cluster.  Their discovery and
characterization by Impey, Bothun and Malin (1988) readily showed 
that the faint end slope of the LF in clusters of galaxies was 
significantly steeper than previous measured.  In the case of Virgo, 
this meant there were galaxies in the range $M_B$ = $-12$ to $-16$ which 
were below the isophotal limits of the plate material used by Sandage.  
Most of these have $\mu_0$ fainter than 24.5 \sb and $\alpha_l$ larger 
than 1 kpc.  The presence of these diffuse galaxies at modest luminosities
increased the faint end power law slope of the LF to a value of $-1.55$;
significantly steeper than the value of $-1.1$ which was thought to 
hold for clusters and perhaps the field (see 
Efthasthiou \etal 1988; Loveday \etal 1992; Marzke \etal 1994).  

\bigskip

The success of the Malin hunt for very diffuse galaxies prompted three new 
surveys.  The first relied on the goodwill of Jim Schombert, who was a
Caltech Postdoc associated with the second Palomar Sky Survey.  Jim was the 
quality control person and thus had direct access to the plates themselves 
for a limited period of time before they were secured in the vault.  This 
allowed an opportunity for hit-and-run visual inspection in pre-defined 
declination strips to search for diffuse galaxies with sizes larger than 
one arcminute.  This produced the catalogs of Schombert and Bothun (1988) 
and Schombert \etal (1992). A second survey was initiated in the Fornax 
cluster using the Malinization technique in order to compare the results 
to Virgo.  These objects are cataloged and described in Bothun \etal (1991)  
which built on the earlier work of Caldwell and Bothun (1987).  The
third survey was initiated with Mike Irwin at Cambridge and this involved
using the Automatic Plate Machine (APM) to scan UK Schmidt plates using an
algorithm optimized to find galaxies of low contrast.  This forms the
most extensive catalog of LSB galaxies to date (Impey \etal 1996).

\bigskip

The goals of these new surveys was to discover, using various techniques,
the extent and nature of this new population of galaxies which had low
contrast with respect to the sky background and hence have remained
undetected and uncataloged.  The importance of discovering this new
population cannot be overstated. The existence of LSB galaxies is a clear signal
that the samples from which we select galaxies for detailed follow-up studies
are incomplete, inadequate and biased.  These surveys are now complete and
most of the results have been published.  They have opened up a new field
of inquiry in extragalactic astronomy.  Detailed studies of the properties
of individual LSB galaxies, and the class as a whole, has resulted in
a number of recent Ph.D. theses:  Knezek (1993), McGaugh (1992), Sprayberry 
(1994), Dalcanton (1995), de Jong (1995), Driver (1995), de Blok (1997),
O'Neil (1997), and Pickering (1997).

\bigskip

In just over a decade, a whole new population of galaxies has been discovered.
Figure 1 provides only a hint on their overall space density.
The number of objects with $\mu_0 \geq 24.0$ \sb is unknown, and can only
be guessed by extrapolation of the trend in Figure 1.  Significant numbers 
of galaxies with $\mu_0 \geq 24.0$\sb have been detected in CCD surveys,
suggesting that the trend remains fairly constant (Dalcanton 1995, O'Neil 
\etal 1997a) or even rises towards fainter $\mu_0$ (Schwartzenberg \etal 
1995). Figure 2 shows an example of one of the more extreme LSB galaxies
turned up in the O'Neil \etal survey.   Reproducing nearly invisible galaxies
on paper is difficult.  Interested readers should inspect the digital
gallery of LSB galaxies available at http://zebu.uoregon.edu/sb2.html.
These LSB galaxies are of cosmological significance and have properties 
which are quite different from those of their HSB
counterparts which dominate existing galaxy catalogs. 

\bigskip
\bigskip
\bigskip
\leftline{\large 4.\quad FORM, PROPERTIES AND STELLAR CONTENT}
\bigskip

In the Hubble Sequence of spirals, morphological classification is based 
on spiral arm texture and definition (see Sandage 1961).  In this case,
classification becomes directly linked to the relative star formation rate
(SFR) via the illumination of the spiral pattern by young stars.  
It is well established that in Hubble Sequence spirals, the bulk of
the star formation occurs within GMC's.  This method of star formation
gives rise to the formation of massive stars within stellar clusters,
which ultimately drives the chemical evolution of galaxies.  Galactic 
disks engage in star formation activity according to a criteria first 
established by Quirk (1972). The more recent expression of this criterion,
by Kennicutt (1989), reflects a convolution of the rotation curve
with the surface density distribution of gas.  The result is a critical 
surface density of H I, as a function of
radius, for the formation of molecular clouds and subsequence massive
star formation.  All data to date (\eg van der Hulst \etal 1993,
de Blok \etal 1996, Pickering \etal 1997) strongly suggests that LSB
disks have an H I distribution which is largely below this threshold.
It is therefore likely that LSB disks represent the
evolutionary track taken by flattened- gas rich systems that are
simply unable to form stars in the familiar environment of molecular clouds.
If correct, this leads to the following set of expectations:

\bigskip

$\bullet$ The arm-interarm contrast in LSB disks should be quite low.

$\bullet$ The overall star formation efficiency should be reduced and
hence LSBs should evolve at a slower rate than those disks that form the
Hubble sequence.

$\bullet$  Massive star formation and hence metal-production should be
inhibited in LSB disks.

$\bullet$ LSBS disks should be relatively low in dust content.

$\bullet$ If the characteristic spacing between newly formed stars 
depends upon the local environment, then star formation may be occurring
in a more diffuse atomic H I medium.   This would produce a lower 
luminosity density and a lower than average surface brightness disk.

$\bullet$ No stars at all should form in LSB disks, if molecular clouds
are truly required to initiate this process in all galactic environments.

\bigskip

The observational properties that have been established for samples of LSB
disk galaxies are largely consistent with these expectations.  The most
salient features are enumerated below:

\bigskip

{\bf 1.} \quad LSB disk galaxies are {\it not} exclusively of low mass.  
This is clearly seen in the distribution of 21-cm velocity widths which is
indistinguishable from any sample of Hubble Sequence spirals, as shown in 
Figure 3.   LSB disks have similar sizes and masses as those high surface 
brightness (HSB) that define the Hubble sequence.  However, despite this 
evidence to the contrary, there is a widespread notion that LSB galaxies 
are exclusively low mass (or ``dwarf'') galaxies.  Indeed, a small percentage 
of the LSB population has truly impressive overall sizes with scale lengths 
that exceed 15 $h_{100}^{-1}$ kpc.  Figure 4 shows that when one closes in 
on a representative sample of galaxies, the entire disk structural plane 
defined by $\alpha_{l}$ and $\mu_0$ is populated.

\bigskip

{\bf 2.} \quad Very few LSB galaxies show evidence for nuclear activity.  
This is in marked contrast to HSB disk galaxies where the percentage which 
have active galactic nuclei can be as high as 50\%, depending on the mean
luminosity of the sample.   The most probable explanation for this 
is that LSB disks generally lack two structural features that facilitate 
gas flows and/or the formation of a compact object in the nucleus -- bulges 
and bars.   In this context it is interesting to note that  about 50\% of 
the large scale length LSB disk galaxies (\eg Malin 1 and cousins) show
some nuclear activity, and all have a normal bulge component with
luminosity $M_B$ = $-18$ to $-20$.  These galaxies are truly enigmatic 
in that ``normal'' formation processes were at work to create the bulge 
component but no conspicuous stellar disk ever formed around this bulge.

\bigskip

{\bf 3.} \quad  The amount of neutral hydrogen in LSB disks is very similar 
to HSB spirals except for their lower than average surface densities.  A 
typical case is shown in Figure 5 where it can be seen that the gas density 
is generally below the threshold for molecular cloud formation and hence
widespread disk star formation and metal production.  LSB spirals are 
significantly deficient in molecular gas compared to HSB spirals of the 
same mass.  Schombert \etal (1990) failed to detect a single LSB disk
in CO (see also Knezek 1993).  The mean upper limit on $M_{H2}$ is an
order of magnitude less than than the $M_{HI} = M_{H2}$ expectation.
In some cases its nearly 2 orders of magnitude less.  A recent calibration
of the H2/CO conversion factor as a function of metallicity (Wilson 1996),
when applied to LSBs changes these limits by only a factor of 3.
Furthermore, virtually none of the LSB galaxies detected by Schombert \etal 
(1992) or Impey \etal (1996) are IRAS sources.  

\bigskip

Although a typical LSB disk contains
a handful of bright H II regions, indicating some current star formation,
the average star formation rate is at least an order of magnitude lower
than in similar mass disk galaxies.  In our galaxy star formation is observed
to only occur in GMCs (\eg Young and Scoville 1991).   LSB galaxies may well 
be devoid of molecular clouds in their interstellar medium but nevertheless 
formed some stars.  We might speculate that the pressure-temperature-density 
manifold in LSB disks physically precludes molecular cloud formation.
The expectation of low metallicity is confirmed by the observations of
McGaugh (1992).   As derived from H II region spectroscopy, the 
abundance of heavy elements such as oxygen, nitrogen, neon and sulfur
can be an order of magnitude lower than HSB disks of the same mass.
On average, LSB disk galaxies with L* luminosities have abundances of $\sim$ 
1/3 solar (McGaugh 1994).    The largest LSBs known (\eg Malin 2, UGC 6614)
may have abundances near solar or in excess of solar - but these are
rare objects.
Thus, LSB galaxies strongly violate the 
mass-metallicity relation defined by Hubble sequence spirals.  This shows
that multiple paths of chemical evolution exist for disk galaxies, which
in turn demands multiple star formation histories.  The LSB disks represent 
some of the most unevolved objects in the universe.  

\bigskip

{\bf 4.} \quad  Practically all of the LSB galaxies discovered to date 
in photographically based surveys (described previously) are blue
despite the obvious lack of star formation.  Mean colors
for these disks range from $U-B = -0.17$, $B-V = 0.49
\pm 0.04$, and $V-I = 0.89$ (McGaugh and Bothun 1994) to 
$B-V = 0.73 \pm 0.05$ and $V-R = 0.50 \pm
0.04$ for the sample of very large scale length LSBs of Sprayberry \etal
(1995a).  In the most extreme cases, these galaxies, despite very low
star formation rates ($\sim$ 0.1 M\solar\ per year), have disk
colors of $B-V \sim 0.1$ and $V-I \sim 0.6$.  By way of comparison with
a large sample of galaxies which obey the Freeman Law,
de Jong \& van der Kruit (1994) found mean colors of  $B-V = 0.75 \pm 0.03$
and $V-R = 0.53$.   One of the more curious aspects of the color
distributions of LSB samples is shown in Figure 6 which indicates 
that there is no correlation between $\mu_o$ and overall color. 
This means that the LSB disks cannot be the faded remnants of HSB
galaxies after star formation has subsided.  

\bigskip

Figure 6 is very difficult to interpret.  Galaxies must fade and redden 
as they age, but we see no evidence of this in our photographically selected
sample of LSBs.   McGaugh (1996) argues that this is a selection effect
which, depending on the plate material, might be
sufficiently severe to prohibit the detection of LSB disks with 
$B-V = 1.0$.   However, we expect that this selection effect is not
that severe as anything with $B-V$ \app 1 and B(0) in the range 23.0 --24.0
\sb should have been detected in these surveys. 
Since we expect that such objects should exist, this suggests 
that the intrinsic local space density of LSB galaxies is even larger than 
suggested by Figure 1.   Perhaps the apparent absence of red LSBs can be
reconciled with a stellar population model and fading scenario which does 
not lead to significant reddening.  One way to offset the tendency for 
stellar populations to redden as they age is to introduce an extremely 
blue horizontal branch which is fed by the evolving stars on the giant 
and asymptotic giant branch.   Such populations have been seen in the 
bulges of spirals and in ellipticals (e.g. O'Connell \etal 1992, Ferguson 
\etal 1991, Brown \etal 1996).  Other hot components such as PAGB stars or PN nuclei would 
also aid in keeping these galaxies bluer as they fade.  In fact, we can 
successfully reproduce the observed colors of LSB disk systems by mixing 
together three populations of stars:

\bigskip

$\bullet$ \quad F2 V stars with $B-V$ = 0.36, $V-I$ = .55 and $M_V$ = 3.3.

$\bullet$ \quad K0 III stars with $B-V$ = 1.00; $V-I$ = 1.30 and $M_V$ = 0.0.

$\bullet$ \quad BHB stars (as in M4) with $B-V$ = $-$0.2; $V-I$ = $-$0.3 and 
$M_V$ = 1.0.

\bigskip

The colors of LSB disks can be reproduced when $\geq$ 50\% of the stars are
F2 and the ratio of hot evolved stars to giants is \app 3:1.  Unfortunately, 
stellar evolution does not predict such a large ratio of  hot evolved stars 
to giants since the giant branch feeds the PAGB stage where the 
lifetime is somewhat shorter.  Hence, this ratio should be
unity at most.  A ratio of unity, however, produce a too red $V-I$ color.
Theoretically, it is also possible that if the hydrogen envelope of 
giant stars is completely lost during their evolution then the remnant core 
will fall on the Helium main sequence which is considerably hotter and more
luminous, at a given mass, than the normal main sequence 
(e.g. an 0.5 M\solar\ Helium main sequence star has $T_{\rm eff}$ \app 45 K).  

\bigskip

The above scenario is admittedly quite extreme, and it cannot be
tested without UV observations of LSB galaxies.  A large program 
aimed at determining fluxes of LSB disks at 2000\AA\ was approved 
as a Guest Investigator program for the UIT Telescope aboard Astro-2.  
Unfortunately,  there was a complete failure in the first stage
image intensifier of the A camera during the Astro-2 mission and none
of the required data could be taken.  Hence, we still don't know if
optically LSB disk galaxies, in fact, have normal surface brightnesses
in the UV.  If they do, there are important implications regarding
deep galaxy surveys (see below).  In fact, a study of a deep Astro-I
image of Fornax (see O'Neil \etal 1996) did show that some optically
LSB objects could be detected at 1500 and 2000 \AA.

\bigskip

If LSB galaxies do not have significant UV fluxes (which also contribute 
to the U and B passbands) then their blue colors are very difficult
to understand.  In most cases, the only physically plausible explanation
is reduced contribution to the integrated light by giants.  This requires
these galaxies to have mean ages several Gyr younger than HSB disks
of the same luminosity. This notion of delayed formation is somewhat 
consistent with
the observed low densities of these objects as the dynamical timescale
(a characteristic timescale over which an object collapses) can be
an order of magnitude larger than HSB galaxies.  Thus we might expect most
LSB galaxies to have late collapse times and hence delayed formation
of their first stars. 

\bigskip

A recent development in the survey and detection of LSB galaxies
can counter the previous tendency to find only blue objects.   The 
first wide field CCD survey specifically designed to detect LSB
galaxies has been initiated by O'Neil \etal (1997a,b).  This survey
was performed with the University of Texas McDonald Observatory
0.8m telescope between 1993 and 1996 using an LF1 2048x2048 pixel CCD
camera with a 1.32'' pixel size.  The sensitivity of this survey
is about 1 \sb fainter than previous photographically based surveys.
Fields in the Pegasus and Cancer clusters (these are Virgo-like 
structures at redshifts 4000-5000 kms$^{-1}$
which are spiral rich) were selected along with many fields in the low
density environment defined by the Great Wall.  The survey yielded the
discovery of 120 new LSB galaxies including, finally, a red component
to the LSB population.  Figure 7 shows the color distributions in
$B-V$ for this sample.  Note again there is no correlation 
between color and $\mu_0$.  Even though this new survey has detected
LSB galaxies with fairly red integrated colors, there is no sign of
a fading sequence.  For this particular sample the average colors are
$B-V = 0.79 \pm 0.04$, $U-B = 0.08 \pm 0.04$, $V-I=0.94 \pm 0.06$.  
Approximately 20\% of the sample has $B-V \geq 0.9$ and $U-B \geq$ 
0.4, the colors that might be expected for a faded, relatively metal-rich 
Hubble sequence spiral.

\bigskip

{\bf 5.} \quad Until very recently, almost nothing was known about the 
dynamics of rotating LSB disks.  Optically, LSB disks galaxies rarely show 
the symmetry that is usually exhibited by Hubble Sequence spirals (but
see Zaritsky and Rix 1997).  To first order, strongly symmetric optical disks
are a signature of rotation dominated kinematics within relatively
cold disks.   The ``chaotic'' optical appearance of many LSB disks
might suggest peculiar kinematics, yet the global 21-cm profiles generally
show the two-horned signature of a rotating disk.    Some LSB disks are
sufficiently gas-rich that their rotation curves can be derived from
aperture-synthesis data.  Two groups (de Blok \etal 1996; Pickering \etal 
1997) have used the VLA and WSRT radio arrays to determine the 2D 
distribution of H I in about a dozen LSB galaxies. The results of 
these efforts can be summarized as follows:

\bigskip

$\bullet$ LSB galaxies have rotation curves that are shallower at small
radii compared to HSB galaxies of the same rotation velocity and mass.

$\bullet$ The gaseous component of LSBs is dynamically significant
at virtually all radii.

$\bullet$ LSB disks are dark matter dominated at virtually all
radii.  Compared to HSB galaxies of the same rotational velocity,
LSB disks have globally and locally higher values of $M/L$ (see also
de Blok \& McGaugh 1996).

$\bullet$ In contrast to HSB galaxies, no ``maximum disk'' mass model
fits the rotation curve.  This means that there is no region in an
LSB disk where the luminous (baryonic) matter dominates the potential
and hence determines the form of the rotation curve.

$\bullet$   Mass models derived from the rotation curves of LSB and 
HSB galaxies show that LSB galaxies inhabit less dense and more extended 
dark matter halos. However, they have dynamical masses comparable to 
those of HSB galaxies.

$\bullet$  The most extreme examples give very hard upper limits to the 
ratio of disk to halo mass. These result in baryon fractions of $\leq$10\% 
at the most, and more likely $\leq$ 3\%.

\bigskip

This dynamical information can be interpreted in terms of the profound 
implication that the density profile of the dark matter halo ultimately 
determines the surface density of the galaxy which forms in that potential.  
Thus, LSB galaxies may well have fundamentally different dark matter 
distributions from HSB galaxies. This makes LSB disks physically distinct 
from HSB disks, even though they may have similar global properties.  The 
paradoxical observation that LSB disks lie on the same Tully-Fisher relation
as HSB disks (see Sprayberry \etal 1995b), albeit with significantly more scatter, can be understood 
if the ratio of dark-to-light matter in disk galaxies is independent
of the dark matter halo density.  In this case, galaxies of the similar
circular velocity but dissimilar surface brightnesses can have the
same luminosity because the LSB galaxy is defined by a larger radius
which contains the mass that determines the circular velocity.  
This implies that LSB disks also have lower surface mass density.

\bigskip

These recent observations of LSBs also give rise to an important new
cosmological inconsistency.  The rotation curves establish that LSB disks 
are everywhere dark matter dominated and reveal a baryonic mass fraction 
($f_b$) of \app 5\% or less.  Another environment which is everywhere dark 
matter dominated is clusters of galaxies.  White \etal (1993) have argued 
that the ratio $\Omega_b/\Omega_o$ measured for a cluster should not be 
significantly different than the universal value.  If this is the case,
then  $\Omega_o$ can be determined from the relation:

$$ \Omega_o = {\Omega_b\over f_b} $$

\noindent where $\Omega_b$ is constrained by the observed abundances of
light elements.
Current observations
indicate that in clusters, $f_b$ $\geq 0.04h^{-3/2}_{100}$.  Combined with the
nucleosynthesis estimate of $\Omega_b \sim 0.015h^{-2}_{100}$ (Walker 
\etal 1991), this leads to $\Omega_o \leq 0.3h^{-1/2}_{100}$.  To 
reconcile this with Einstein-DeSitter models requires either $H_0$ $\leq$ 
30 kms$^{-1}$ Mpc$^{-1}$, or that total cluster masses have been 
systematically underestimated.  The latter possibility does not appear to  
be the case (Evrard \etal 1996) and hence the measured values of $f_b$
in clusters appears quite inconsistent with $\Omega = 1$.  In contrast,
$f_b$ as measured in LSB disks is significantly lower than the cluster
value and is consistent with $\Omega = 1$.  So which is the more
representative baryonic repository -- clusters of galaxies or LSB
disks?  Arguments given in Impey and Bothun (1997) show that LSB galaxies
make a significant contribution to the total baryonic mass density 
in the Universe, providing up to half of it depending on the faint
end slope of the luminosity function (see below).  

\bigskip
\bigskip
\bigskip
\leftline{\large 5.\quad COSMOLOGICAL RELEVANCE OF LSB GALAXIES}
\bigskip

The discovery of a substantial population of LSB galaxies at z = 0 
is relevant to a number of cosmological issues.  These issues are
enumerated below; those issues that are only briefly discussed here
are covered in more detail in Impey and Bothun (1997).

\bigskip

{\it QSO absorption lines:} \quad  Increasing the number density of 
galaxies is a very good first step towards understanding the origin of 
QSO absorption lines.  In addition, the discovery of Malin 1, which 
has 10$^5$ kpc$^2$ of gaseous cross section at a level of $N_h \geq
3 \times 10^{19}$ cm$^{-2}$ is a helpful addition to the zoology 
of objects that might produce damped Ly$\alpha$ systems. 

\bigskip

{\it Large Scale Structure (LSS):} \quad To date, most studies of 
LSS are based on redshift surveys of optically selected samples, 
which, by definition, do not contain many LSB galaxies.  If LSB 
galaxies are better tracers of the mass distribution, then
the linear biasing factor between mass and light has a surface brightness
dependence.  This would be an ugly complication to the various
sophisticated attempts to determine $b\Omega^{0.6}$,
where $b$ is the (scale-independent) linear bias parameter 
(see Strauss 
\& Willick 1995).  In an attempt to shed light on this, Bothun 
\etal (1985,1986) performed a redshift survey of \app 400 LSB disk 
galaxies, primarily selected from the UGC.  Few of these galaxies 
were in the CFA redshift survey at that time.  That data, coupled 
with a thorough analysis of the clustering properties of LSB galaxies
 by Bothun \etal (1993) and Mo \etal (1994) provides the following 
insights:

\bigskip

$\bullet$ On scales $\geq 5 h^{-1}_{100}$ Mpc LSB galaxies trace the 
identical structure as HSB galaxies.   

$\bullet$ LSB galaxies generally avoid virialized regions.

$\bullet$ On scales $\leq 2 h^{-1}_{100}$ Mpc LSB galaxies are significantly
less clustered.  In fact, there is a significant deficit of companions
within a projected distance of $0.5 h^{-1}_{100}$ Mpc indicating that 
LSB disks are generally isolated.  In general, the correlation function
has a lower amplitude for LSBs (see Mo \etal 1994) resulting in them
being less clustered on all scales, compared to the HSB galaxies contained
in the CFA redshift survey.

$\bullet$ Whether in  groups or clusters, LSB galaxies are usually near 
the edge of the galaxy distribution.  The results of O'Neil \etal (1997a)
show that, in the cluster environment, there is a limiting value of
$\mu_0$ which depends on local galaxy density.  If one considers that
the highest surface brightness galaxies of all, ellipticals, tend to
occur in virialized regions then we have the surface brightness extension
of the morphology-density relation of Dressler (1980).  That is, the
local galaxy density limits the surface mass/luminosity density of
galaxies.

$\bullet$ LSB galaxies are not  preferentially found in large scale
voids.  For instance, there are none in either the CFA bubble (Bothun \etal
1992) or the Bootes Void (Aldering \etal 1997).

\bigskip

The following scenario can be advanced.  Given a Gaussian initial 
spectrum of density perturbations which will eventually form galaxies, 
there should be many more low density perturbations than high density 
ones.  Many of these low density perturbations are subject to disruption or
assimilation into other perturbations and hence will not produce individual 
galaxies.  However, if a substantial percentage of them do survive to 
produce individual galaxies then we would expect them to dominate the 
galaxy population.    We tentatively identify the LSB galaxy population
with these 1-2$\sigma$ peaks in the initial Gaussian perturbation
spectrum.  If that is the case, we expect LSB galaxies to (1) have
formed in isolation, and (2) be fair tracers of the mass distribution 
on large scales.  This isolation on small scales must clearly affect
their evolution since, compared to HSB galaxies, LSB galaxies have 
experienced fewer tidal encounters with nearby galaxies over a Hubble 
time.  Tidal encounters are effective at clumping gas and driving
global star formation.  Without this external hammer, LSB galaxies 
would continue to evolve slowly and passively.  

\bigskip

{\it The Faint End Slope of the Galaxy Luminosity Function:} \quad  
In recent years, a lot of different values have appeared for the
faint end slope, $\alpha$, of the Schechter luminosity function.
The canonical value of $\alpha$ \app $-$1.25 for HSB field galaxies
can be compared with observations of nearby clusters that show $\alpha$ 
of \app $-$1.1 (but see Wilson \etal 1997 for significantly steeper
values for more distant clusters).  However, $\alpha$ = -1.9 
has been determined for low mass irregular
galaxies in the CFA redshift survey (Marzke \etal 1994).   The safest
thing to then conclude is that recent surveys have now produced a variety
of different values of $\alpha$ indicating that its a) uncertain and
b) may evolve with time.

\bigskip
In general, corrections for incompleteness to the galaxy luminosity
function have been erroneously applied in the past based on apparent 
magnitude only.  This is conceptually incorrect since galaxies are 
selected on the basis of a combination of surface brightness and 
luminosity. This is best illustrated in the Virgo and Fornax clusters,
where the diffuse and fairly large LSB dwarf were missed in earlier 
surveys but detected using the Malin method.  The integrated luminosity 
of these galaxies is brighter then many of the Binggeli \etal (1984)
dwarfs due to their large scale length.  The surveys in Virgo (Impey 
\etal 1988) and Fornax (Bothun \etal 1991) revealed LSB galaxies that 
were up to 3 magnitudes  brighter than the nominal magnitude limit of 
the Binggeli \etal (1984) survey but which were missed by that survey.  

\bigskip
Figure 8 (adapted from Figure 10 in Impey \etal 1988) illustrates
the basic point for the Virgo cluster.  
When these "missing galaxies" are properly accounted for, there is 
a significant increase in the faint end slope of the luminosity function
relative to that which is obtained simply by applying a correction
for incompleteness based on apparent flux.  The correlation between
surface brightness and magnitude that seems to hold for dwarf
elliptical galaxies (\eg Sandage \etal 1985;
Binggeli \etal 1985; Caldwell and Bothun 1987;
Bothun \etal 1989) has been shown to break down below $\mu_o$ \app
24.5 (\eg Bothun \etal 1991) 
indicating that corrections for apparent flux do not mimic the
more proper correction based on surface brightness selection effects
(see also Sprayberry \etal 1997; Impey and Bothun 1997).  
Combining the Virgo and Fornax cluster sample of dwarfs yields $\alpha$ = 
$-$1.55 $\pm$ 0.05 after these proper corrections are made.
Such a steepening indicates that LSB galaxies dominate numerically in 
clusters.  If stellar/baryonic
$M/L$ increases with decreasing surface brightness for this 
cluster population then they contain a significant fraction of the baryonic 
material in clusters (which might exacerbate the $f_b$ problem in clusters).
It is interesting to speculate that cluster LSB galaxies might represent
the results of a phase of intense baryonic blow-out at higher redshifts.  
These relic galaxies then might be the source of the gas seen in clusters 
as well as the enrichment of that gas.  In the case of the Virgo cluster, 
there are approximately twice as many dwarfs as brighter galaxies.

\bigskip

The APM survey of Impey \etal (1996) allows the contribution of LSB
galaxies to the field galaxy luminosity function to be estimated.
This has been done in detail by Sprayberry \etal (1997) who find 
a faint end slope of $\alpha$ = $-$1.42 compared to $\alpha$ = $-$1.09
for all disk types in the CFA redshift survey.   It is too soon to
tell if there is a significant difference between the cluster and
field faint end slopes but in both cases, the discovery of LSB 
galaxies in both environments yields the same result: the faint
end slope steepens. 

\bigskip

{\it The Faint Blue Galaxy Connection:} \quad   One immediate result 
from various deep surveys of galaxies (\eg Lilly \etal 1991,  Lin \etal 1997,
Odewahn \etal 1996,  Driver \etal 1996, Glazebrook \etal 1995) was 
the discovery  of an apparent excess, relative to local observations,
of faint galaxies with blue colors.  These enigmatic faint blue galaxies 
(FBGs) are the subject of much research.  Since the FBG population 
apparently does not exist at z = 0, reasons for their absence from 
galaxy catalogs of the local universe must be found. Two possible 
approaches involve either a very large merger rate since z \app 0.7 
or rapid fading of this FBG population (Efstathiou \etal 1991; Colless 
\etal 1990).  This faded population is potentially an important constituent 
of the total baryonic content of the universe (see Cowie 1991; Bouwens
and Silk 1996).  However, 
arguments presented above suggest that these faded remnants have not yet 
been detected in large numbers (see also Dalcanton 1993).

\bigskip

A number of explanations for the FBG ``problem'' have a natural linkage
with LSB galaxies at z = 0.  Some of the possible explanations are the 
following:

\bigskip

$\bullet$ If optically selected
LSB galaxies do have normal UV fluxes, then this population can become
conspicuous in faint galaxy surveys  when this light is redshifted
into the ground based filter set.  This would occur, in general, over
the redshift range 0.5-1.  It is possible that some of the FBGs
are in fact LSB galaxies!   For example, the median luminosity of the 
FBGs in Lilly \etal (1991) is identical to the LSB sample of McGaugh \&
Bothun (1994) and both samples are dominated by late type galaxies
which are weakly clustered.

\bigskip

$\bullet$ The FBGs are a population of star bursting dwarf galaxies
located at modest redshift.  This suggestion takes advantage of the
fact that in any GLF with $\alpha \leq -1$, low mass dwarf galaxies
dominate the space density.  To produce the FBGs, however, these
dwarf galaxies have to be at least an order of magnitude brighter
at modest redshifts which requires a fairly significant star
formation rate.  Subsequent heating of the ISM by massive stars
and supernova should be sufficient to heat it beyond the escape
velocity of these low mass systems (see Silk \etal 1987;
Wyse and Gilmore 1992) and hence
such galaxies have a significant phase of baryonic blowout after
which they fade to very low absolute luminosities and hard to
detect at z=0.  This mechanism gives the universe a channel for 
making baryons ``disappear'' with time.  This hypothesis has been
investigated in detail by Phillips and Driver (1995) and Babul
and Ferguson (1996).

\bigskip

$\bullet$ The number density of galaxies is not conserved and the
FBGs merge with other galaxies.  It is difficult to support this
hypothesis because (1) the FGBs are already weakly clustered, and
(2) the required merging rate is significantly higher than the
inferred rate at modest redshift by Patton \etal (1997).  The
merger idea works best if the FBGs are predominately at higher
redshift, where the merger rate is higher owing to the smaller
volume of the universe.

\bigskip

$\bullet$ The FGBs represent an entirely new population of galaxies --
one defined by a star formation history or an initial mass function
that allows only a limited window of visibility before the galaxies
fade to extremely low surface brightness levels by z=0.  As an example,
an IMF which forms no stars less than 1 M\solar produces
a galaxy that is destined to have a very bright phase and then
fade into mostly white dwarf remnants at z = 0.  A collection of 
10$^{11}$ white dwarfs distributed in a disk with scale length of
2 kpc would have a central surface brightness of $\mu_0$ = 27.5 
which would be below the current detection threshold of LSB galaxies.
This situation illustrates the kind of extreme fading which would
be necessary to hide this component from being detected locally.
Interestingly, the latest MACHO results (see Sutherland \etal 1996) 
strongly suggest that halo white dwarfs are the main cause of the
8 observed lensing events towards the LMC.  In that sense, the Galactic
halo might be regarded as a LSB galaxy.

\bigskip

$\bullet$ The FBGs are an artifact of uncertainties in the determination
of the local galaxy luminosity function. In particular, the faint end 
slope has been seriously underestimated from nearby samples.  Alternatively,
the local normalization, $\Phi(0)$, of the galaxy luminosity function could 
be too low.  This would result if, for instance, deep surveys were more 
efficient at selecting galaxies than nearby surveys.   Driver \etal 
(1994a,b) have explored this possibility and concluded that the local 
normalization may well be seriously underestimated.   However, the effect 
of increasing the space density at z = 0 can only partially offset the 
excess FBG counts.  A much larger lever arm is provided by steepening 
$\alpha$.

\bigskip

Sprayberry \etal (1997) have considered the effects of increasing
$\alpha$ (due to LSB inclusion) and increasing $\Phi(0)$ on the
ability for LSB galaxies to explain the FBG excess.  The conclude
that the CFA survey has missed about 1/3 of the total galaxy population
at z = 0.  While this helps to resolve the apparent difference between
the large numbers of FBGs and the local galaxy population, this effect
by itself is not large enough to completely resolve the issue (see also 
Ferguson and McGaugh 1994).  For additional assistance we turn to the 
Lilly \etal (1995) redshift survey of \app 500 faint galaxies.  Their 
sample has excellent quality control and is fairly free from selection 
effects and is primarily aimed at determining the galaxy luminosity 
function up to a redshift \app 1.

\bigskip

Lilly \etal detect a change in the luminosity function of blue galaxies
by approximately one magnitude between z \app 0.38 and z \app 0.62, and 
another magnitude between z \app 0.62 and z \app 0.85.  Moreover, many 
of these galaxies have been observed with HST in order to measure
characteristic surface brightnesses.  Schade \etal (1995) find that
the disks of these blue galaxies are \app 1 magnitude higher in
surface brightness at z = 0.8 than z = 0.3.  These studies provide
rather strong evidence for luminosity evolution in the FBGs.
In a 15 Gyr old universe, there are approximately 3.3 billion years
between z = 0.85 and z = 0.38.  The data indicate that a typical
FBG would decline in luminosity by a factor of six (\eg 2 mags) over this 
time period.  This modest decline is quite consistent with
standard population synthesis models involving a normal IMF which fades
after a significant burst of star formation has occurred to make
these objects visible at higher redshift.
The decline in luminosity is primarily a reflection of the disappearance
of the upper main sequence.  By z = 0, these galaxies will certainly
not have faded to levels that preclude their detection, and indeed
many of them are likely to be represented by the z = 0 red LSB population
which has now been discovered.

\bigskip
\bigskip
\bigskip
\leftline{\large 6. \quad SUMMARY}
\bigskip

We conclude this review by again referring to Figure 1. These results on 
the space density of galaxies as a function of central surface brightness 
require a basic adjustment in the way we think about galaxies.  Much of 
the current thought is implicitly one dimensional, with one parameter 
(like luminosity or morphology) dominating the way problems are approached.  
This is  no longer sufficient. Surface brightness selection effects have 
been severe.  Our results now indicate that up to 50\% of the general 
galaxy population is in the form of galaxies with central surface 
brightness below 22.0 \sb.  Moreover, the space density remains flat 
out to the limits of the data, and the space density of the lowest
surface brightness disks ($\mu_0$ \app 25.0 \sb, the limits of current
data) is vastly higher than would have been anticipated based on 
Freeman's Law.  

\bigskip

LSB galaxies offer a new window onto galaxy evolution.  Because of this,
the quest to find LSB galaxies continues.  Over the next few years we hope 
to extend our sensitivity by using wide field CCD surveys of the sky at 
dark sites.  These and new surveys by other groups should extend the 
current data by two magnitudes down to $\mu_0$ = 27.0 \sb.   The major 
goals of these new surveys are to determine if the space density of 
galaxies as a function of $\mu_0$ continues to remain flat over a factor 
of 100 in $\mu_0$, and to detect what we so far have failed to detect 
in large numbers -- the red LSB population that must result from the 
faded remnants of galaxies that no longer can form stars.  In this regard, 
the Sloan Digital Sky Survey will be extremely helpful if its automatic 
image recognition system can reliably detect LSB galaxies.  At levels
well below the sky brightness, however, LSB galaxies are often defined by 
disconnected regions of pixel intensity.  While the human eye is 
rather good at finding these, its not clear if that particular 
algorithm can be reproduced in a machine.

\bigskip

Despite the power of CCD surveys and automated detection algorithms, 
we are inevitably brought back to the starting point of this review.  
Messier cataloged galaxies long before their extragalactic nature was 
understood, and it is ironic that his catalog was a reject list of 
stationary objects for comet-hunters to avoid.   We now know a lot more 
about about galaxies, but the selection biases that have operated for
200 years have not been fully overcome. It would be unwise to presume
that we have yet revealed the true population of galaxies.

\bigskip

We leave the reader with a small list of things that we hoped they
have learned from this review article:

\bigskip

1.  LSB galaxies exist.  Lurking beneath the brightness of the night
sky are real galaxies with evolutionary histories substantially 
different from the processes that produced the Hubble Sequence of 
spirals.  In particular, LSBs evolve at a significantly slower rate 
and may well experience star formation outside of the molecular 
cloud environment.

\bigskip

2.  Surface brightness selection effects have been severe.  A proper
accounting of them has increased the local number density of galaxies
and steepened the faint end slope of the galaxy luminosity function.
Despite this progress, these selection effects still exist and thus
we do not yet have a representative, volume limited sample of nearby
galaxies.

\bigskip

3.  LSB galaxies span the entire galactic mass range.  They are not
exclusively low mass galaxies but include the most massive disk
galaxies discovered to date (\eg Malin 1).  LSB disks likely are the 
manifestation of 1-2$\sigma$ isolated peaks in the initial density 
fluctuation spectrum.  These lower density peaks have longer collapse 
times and trace the mass distribution in a relatively unbiased way.

\bigskip

4.  LSB galaxies are embedded in dark matter halos which are of lower
density and more extended than HSB galaxy halos.  In this sense, disk
galaxy surface mass density and subsequent evolution may be predetermined
by the form of the dark matter halo.  Surface mass density appears to be
the single biggest driver of disk galaxy evolution.

\bigskip
\bigskip

ACKNOWLEDGMENTS
\bigskip
\bigskip
\bigskip

A number of people have helped to support this project over the
last decade.  We gratefully acknowledge Mike Disney for making
us think, David Malin for his wizardry
and patience with us, Jim Schombert for being there, Steve Strom
for pointing the way, Jay Gallagher for telling one of us (GDB) to
work on something \lq\lq hard",  Mark Cornell for assistance with
the Texas observations, and Allan Sandage for originally showcasing
the smudge galaxies.  We gratefully acknowledge support from the NSF under
grant AST-9005115 and AST-9003158 without which, this project could
never have been sustained.

\vfil\eject

\centerline{\bf References}
 
\bigskip
\bigskip

Aldering, G., Bothun, G., Kirshner, R., and Marzke, R. 1997 {\sl Ap. J.}
submitted

Babul, A., and Ferguson, H. 1996 {\sl Ap.J.} 458:100

Binggeli B, Sandage A, Tarenghi M. 1984. {\sl Astron. J.} 89:64

Binggeli B, Sandage A, Tammann GA. 1985. {\sl Astron. J.} 90:1681 

Boroson T. 1981. {\sl Ap. J. Suppl.} 46:177

Bothun, G. 1981,  {\sl Ph.D. Thesis.} University of Washington

Bothun, G., Beers, T., Mould, J., and Huchra, J. 1985 {\sl Astron. J.} 90:2487

Bothun, G., Beers, T., Mould, J., and Huchra, J. 1986 {\sl Ap.J.} 308:510

Bothun, G., Caldwell, N., and Schombert, J. 1989 {\sl Astron. J.} 98:1542

Bothun, G., Geller, M., Kurtz, M., Huchra, J., and Schild, R. 1992

{\sl Ap.J.} 395:349

Bothun, G., Schombert, J., Impey, C., Sprayberry, D., and McGaugh, S. 1993

{\sl Astron. J.} 106:530

Bothun GD, Impey CD, Malin DF, Mould JR. 1987. {\sl Astron. J.} 94:23

Bothun GD, Impey CD, Malin DF. 1991. {\sl Ap. J.} 376:404

Bouwens, R. and Silk, J. 1996 {\sl Ap. J. Letters} 471:L19

Brown, T., Ferguson, H., and Davidsen, A. 1996 {\sl Ap.J.} 472:327

Caldwell N, Bothun GD. 1987. {\sl Astron. J.} 94:1126

Colless, M., Ellis, R., Taylor, K., and Hook, R. 1990 {\sl MNRAS} 244:408

Cowie, L. 1991 {\sl Physica Scripta} 36:102

Dalcanton, J.J. 1993, ApJ, 415, L87

Dalcanton JJ. 1995. {\sl Ph.D. Thesis.} University of Princeton

Dalcanton, J.J., Spergel, D.N., \& Summers, F.J. 1997, {\sl Ap. J.} in press

Davies, J. 1990 {\sl MNRAS} 244, 8

de Blok, E. 1997  {\sl Ph.D. Thesis.} University of Groningen

de Blok, E., and McGaugh, S. 1996 {\sl Ap. J. Letters} 469, L89

de Blok, E.,  McGaugh, S. and van der Hulst, T. 1996 MNRAS 283, 18

de Jong RS. 1995. {\sl Ph.D. Thesis.} University of Groningen

de Jong, R.S. 1996. {\sl A\&A} 313, 45

de Jong, R.S., \& van der Kruit, P.C. 1994, A\&AS, 106, 451

Disney MJ. 1976. {\sl Nature} 263:573

Disney MJ, Phillips S. 1983. {\sl MNRAS} 205:1253

Dressler, A. 1980 {\sl Ap.J.} 236:351

Driver SP, Phillips S, Davies JI, Morgan I, Disney MJ. 1994a. {\sl MNRAS} 268:393

Driver, S.P., Phillips, S., Davies, J.I., Morgan, I., \& Disney, M.J. 

1994b, MNRAS, 266, 155

Driver, S.P., 1995  {\sl Ph.D. Thesis.} University of Cardiff

Driver, S., Couch, W., Phillips, S., and Windhorst, R. 1996 {\sl Ap. J. Letters}

466, L5

Efstathiou G, Ellis RS, Peterson BA. 1988. {\sl MNRAS} 232:431 

Efstathiou, G., Bernstein, G., Tyson, J.A., Katz, N., \& Guhathakurta, P. 

1991, ApJ, 380, L47

Evrard, A., Metzler, C., and Navarro, J. 1997 {\sl Ap.J.} 469:494

Ferguson, H., \etal 1991 {\sl Ap.J. Letters} 382, L69

Ferguson HC, McGaugh SS. 1995. {\sl Ap.J.} 440:470

Freeman KC. 1970. {\sl Ap. J.} 160:811

Glazebrook, K., Ellis, R., Colless, M., Broadhurst, T.,
Allington- Smith,

J., and Tanvir, N. 1995 {\sl MNRAS} 273:157

Hubble, E. 1922 {\sl Ap. J.} 56:162

Impey CD, Bothun GD. 1989. {\sl Ap.J} 341:89

Impey CD, Bothun GD, Malin DF. 1988. {\sl Ap. J.} 330:634

Impey CD, Sprayberry D, Irwin MJ, Bothun GD. 1996. 

{\sl Ap. J. Suppl.} 105:209

Impey, CD. and Bothun, GD. {\sl Ann. Rev. Astr. Ap.} in press

Kennicutt RC. 1989. {\sl Ap.J.} 344:685

Knezek PM. 1993. {\sl Ph.D. Thesis.} University of Massachusetts

Kormendy J. 1977. {\sl Ap.J.} 217:406

Lilly, S.J., Cowie, L.L., \& Gardner, J.P. 1991, ApJ, 369, 79

Lilly, S.J., Tresse, L., Hammer, F., Crampton, D., and Lefevre, O.

1995 {\sl Ap.J.} 455:108

Loveday J, Peterson BA, Efstathiou G. Maddox SJ. 1992. {\sl Ap. J.} 390:338

Marzke RO, Geller MJ, Huchra JP, Corwin Jr HG. 1994. 

{\sl Astron. J.} 108:437

McGaugh SS. 1992. {\sl Ph.D. Thesis.} University of Michigan

McGaugh SS. 1994. {\sl Ap. J.} 426, 135

McGaugh, S.S. \& Bothun, G.D.  1994, AJ, 107, 530

McGaugh SS, Bothun GD, Schombert JM. 1995. {\sl Astron. J.} 110:573

McGaugh SS. 1996. {\sl MNRAS} 280:337 

Mo HJ, McGaugh SS, Bothun GD. 1993. {\sl MNRAS} 267:129

Nilson PN. 1973. {\sl Uppsala General Catalog of Galaxies.} Uppsala, 

Finland: Uppsala Astronomical Observatory

O'Connell, R., \etal 1992 {\sl Ap.J. Letters} 395:L45

Odewahn, S., Windhorst, R., Driver, S., and Keel, W. 1996 

{\sl Ap. J.  Letters} 472, 13

O'Neil, K., Bothun, G., Smith, E.P., and Stecher, T. 1996 

{\sl Astron. J.} 112:431

O'Neil, K. 1997  {\sl Ph.D. Thesis.} University of Oregon

O'Neil, K., Bothun, G., and Cornell, M. 1997a, {\sl Astron. J.} in press

O'Neil, K., Bothun, G., Cornell, M and Impey, C. 1997b 

{\sl Astron. J.} submitted

Patton, D., Pritchet, C., Yee, H., Ellingson, E., and Carlberg, R.

1997 {\sl Ap.J.} 475:29

Phillips, S., Disney, M., Kibblewhite, E., and Cawson, M.

1987,  MNRAS 229, 505

Phillips, S., \& Driver, S.P.  1995, MNRAS 274,83

Pickering, T. 1997  {\sl Ph.D. Thesis.} University of Arizona

Pickering, T., van Gorkom, J., Impey, C., and Bothun, G. 1997

{\sl Astron. J.} submitted

Quirk, W.J. 1972, {\sl Ap. J. Letters} 176,L9 

Romanishin, W., Strom, K.M., \& Strom, S.E. 1983, ApJ, 263, 94 

Sandage, A. 1961 {\sl The Hubble Atlas of Galaxies}  Carnegie

Institute of Washington, Pub No. 618

Sandage A, Binggeli B, Tammann GA. 1985. {\sl Astron. J.} 90:1759

Schade, D., Lilly, S., Crampton, D., Hammer, F., LeFevre, O., and

Tresse, L. 1995 {\sl Ap.J. Letters} 451:L1

Schombert JM, Bothun GD. 1988. {\sl Astron. J.} 95:1389

Schombert JM, Bothun, GD, Impey, CD, and Mundy, L. 1990 

{\sl Astron. J.} 100:1523

Schombert JM. Bothun GD, Schneider SE, McGaugh SS. 1992. 

{\sl Astron. J.} 103:1107

Schwartzenberg JM, Phillips S, Smith RM, Couch WJ, Boyle BJ. 

1995 {\sl MNRAS} 275:171

Silk, J., Wyse, R., and Shields, G. 1987 {\sl Ap. J. Letters} 322:L59

Sprayberry, D. 1994  {\sl Ph.D. Thesis.} University of Arizona

Sprayberry D, Impey CD, Bothun GD, Irwin MJ. 1995a 

{\sl Astron. J.} 109:558

Sprayberry, D., Bernstein, G.M., Impey, C.D., \& Bothun, G.D. 1995b,

ApJ, 438, 72

Sprayberry D, Impey CD, Irwin MJ, Bothun GD. 1997 {\sl Ap. J.} in press

Strauss, M., and Willick, J. 1995 {\sl Physics Reports} 261:271

Sutherland, W., \etal 1996  To appear in
proceedings of workshop

on ``Identification of Dark Matter'', Sheffield, Sep. 1996 

van der Hulst JM, Skillman, ED, Smith TR, Bothun GD, McGaugh SS, 

and de Blok, WJG.  1993. {\sl Astron. J.} 106:548

Walker TP, Steigman G, Schramm DN, Olive KA, Kang H-S. 1991. 

{\sl Ap.J.} 376:51

White SDM, Navarro JF, Evrard AE, Frenk CS. 1993. {\sl Nature} 366:429

Wilson, C. 1996 preprint

Wilson, G., Smail, I., Ellis, R., and Couch, W. 1997 {\sl MNRAS} in
press

Wyse, R. and Gilmore, G. 1992 {\sl Astron. J.} 104, 144

Young, J. and Scoville, N., 1991 {\sl Ann. Rev. Astr. Ap} 29:581

Zaritsky, D. 1997 {\sl Ap. J.} in press

Zaritsky, D. and Rix, H. 1997 {\sl Ap. J.} 477, 118

Zwicky F. 1957. In {\sl Morphological Astronomy.} New York: 

Springer-Verlag

\clearpage

\centerline{\bf Figure Captions}

\bigskip
\bigskip

{\bf Figure 1:} The space density of galaxies as a function of central
surface brightness.  LSB objects appear to the left in this diagram.
Raw counts from the indicated surveys have been converted to space
density through the use of volumetric corrections discussed here and
in more detail in McGaugh \etal 1995.  The solid line shows the
surface brightness distribution which Freeman's Law suggests.  The flat
line fit to the data, from McGaugh 1996,  has a space density which is
6 orders of magnitude higher than predicted from Freeman's Law.

{\bf Figure 2:}  An example from the Texas survey (\eg O'Neil \etal 1997a)
of an extremely diffuse object.  The image size is 1.5x1.5 arc minutes
and the galaxy looks to be at least one arc minute in "diameter" at
a probably redshift of 4-6,000 km/s.  A better presentation of these
diffuse galaxies can be found at http://zebu.uoregon.edu/sb2.html

{\bf Figure 3:} Fractional distribution of 21-cm velocity widths for
samples of LSB and HSB galaxies.   The data are 50\% velocity widths.
The LSB sample represents 131 galaxies which exhibit double-horned
profiles from the Schombert \etal (1992) catalog and the HSB represents
1500 galaxies with double-horned profiles whose data has been compiled
in the literature.

{\bf Figure 4:} Plot of $\mu_o$ vs $\alpha_l$ for samples of LSB
and HSB galaxies.  The absence of HSB galaxies with large values of
$\alpha_l$ is real; other than that, the two variables are uncorrelated.
The LSB data come from McGaugh and Bothun (1994) and Sprayberry (1994)
and the HSB data come from de Jong (1995).

{\bf Figure 5:}  An example of a LSB disk galaxy rotation curve and
surface H I distribution plotted against the critical surface density
criterion of Kennicutt (1989).  The galaxy depicted here is UGC 6614
where the data come from Pickering \etal (1997).  The surface density
of H I is everywhere below the critical density.  At a radius of 30 kpc
it rises to meet the critical density and a small amount of star
formation, as evidenced by weak H II regions, is observed there (see
data in McGaugh 1994).  Most all LSB galaxies have measured surface
densities of H I which are below the critical density (see also 
van der Hulst \etal 1993; de Blok \etal 1996).
 
{\bf Figure 6:}  Non-correlation between the integrated disk color and
observed $\mu_o$ for the LSB sample of McGaugh and Bothun (1994).  There
is clearly no trend of increasing $B-V$ with decreasing  $\mu_o$ as
would be expected for a fading scenario.  If anything, the mean $B-V$
color get slightly bluer with decreasing  $\mu_o$.

{\bf Figure 7:} Plot of observed $\mu_o$ and  $B-V$ for the Texas
CCD based sample of O'Neil \etal (1997a).  While there is still no
correlation between $\mu_o$ and  $B-V$ , the percentage of objects with
$B-V$ $\geq$ 0.8 is significantly higher in this sample compared to
samples of LSB galaxies which are photographically selected.  Consistent,
with those previous samples, however, very blue and very low $\mu_o$
galaxies also are present in this sample.

{\bf Figure 8:}  The luminosity function of the Virgo cluster galaxies
showing the difference between corrections based on apparent magnitude
from those based on surface brightness selection effects.  The solid
circles/squares give the raw counts and error bars as a function of
apparent magnitude.  The filled circles are the result of applying
an incompleteness correction based only on apparent magnitude.  The
open circles are the result of applying a correction based on 
surface brightness.   These galaxies are missed not because of
reduced apparent flux, but because too much of their flux is below
the night sky background.    As a result, the correction correction
for incompleteness based on surface brightness considerations sets
in at a substantially brighter magnitude than corrections based only
on apparent flux.  These are the galaxies in Virgo that were discovered
by Impey \etal 1988.

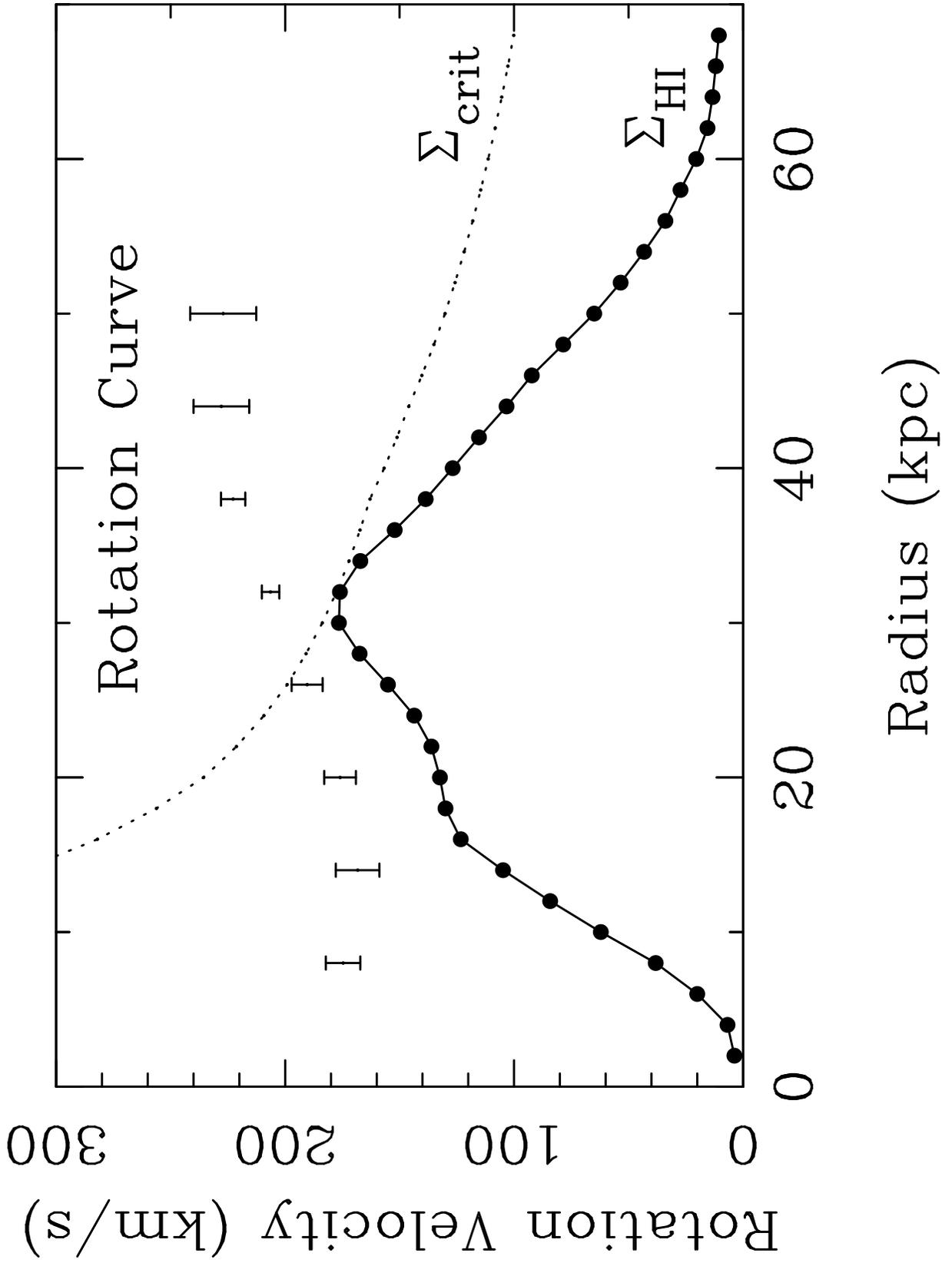
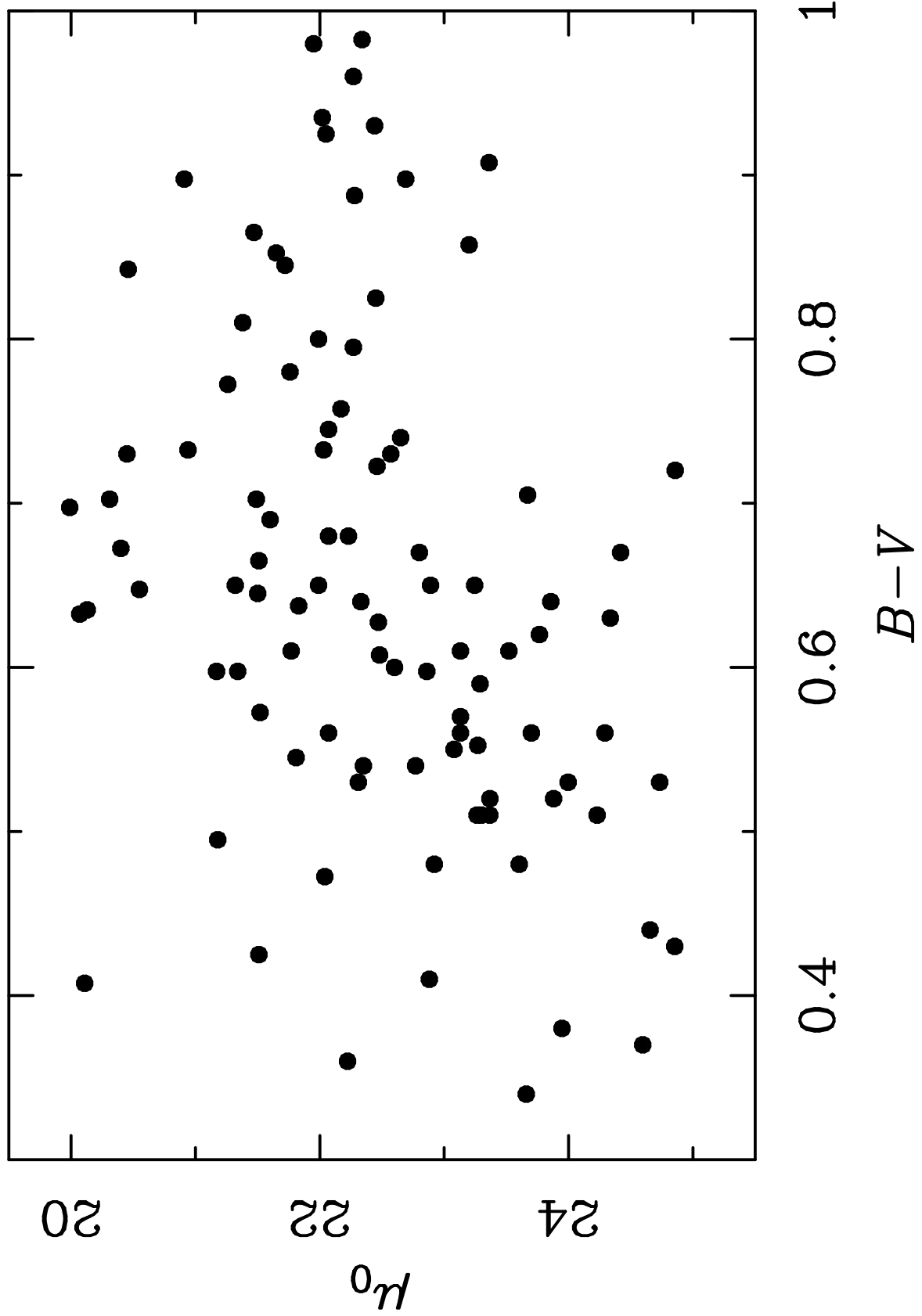
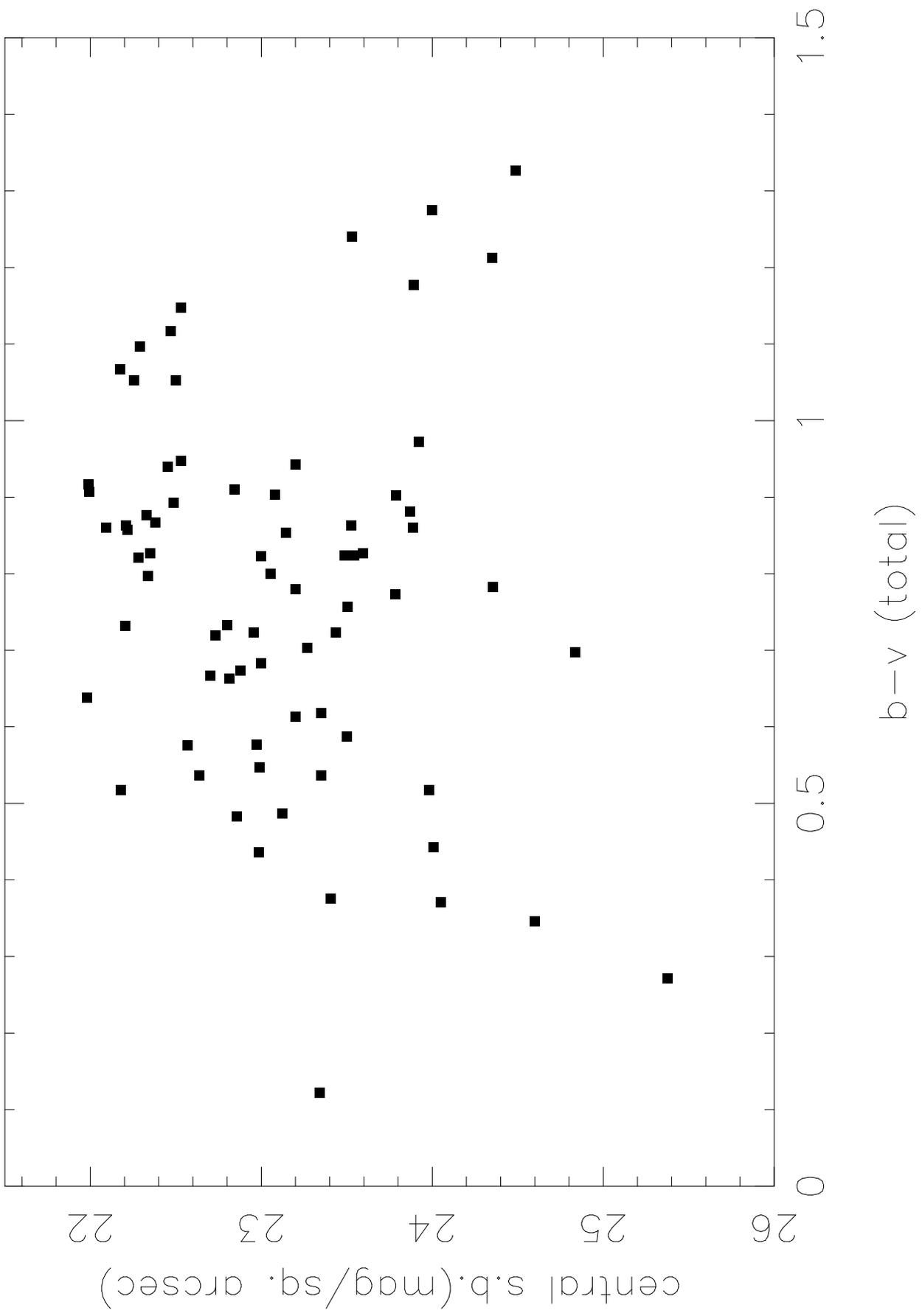
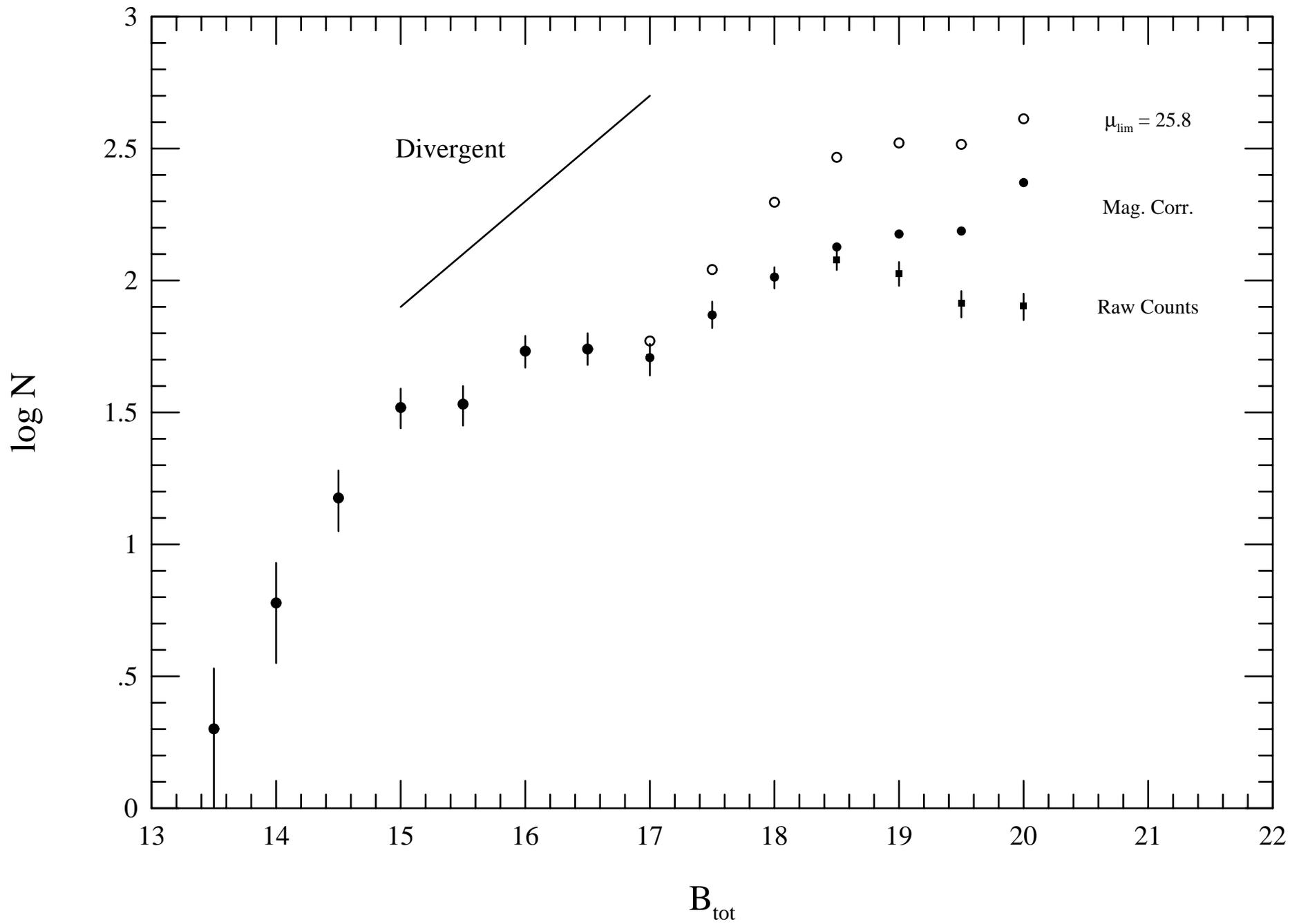
\end{document}